\documentclass[preprint,12pt]{elsarticle}
\journal{}
\usepackage{color}

\begin{document}

%\begin{frontmatter}

\title{Dynamical Casimir effect in a Kerr Cavity}

\author{R. Rom\'an-Ancheyta\corref{mycorrespondingauthor}}
\cortext[mycorrespondingauthor]{Corresponding author}
\ead{ancheyta6@gmail.com}
\author{C. Gonz\'alez-Guti\'errez}
\ead{carlosgg04@gmail.com}
\author{J. R\'ecamier}
\ead{pepe@fis.unam.mx}
\address{Instituto de Ciencias F\'{\i}sicas, 
Universidad Nacional Aut\'onoma de M\'exico, \\C.P. 62210 Cuernavaca, Morelos, M\'exico}

\date{16 Mar 2016}

%\end{frontmatter}

\begin{abstract} 
We study the dynamical Casimir effect  in an electromagnetic cavity containing a Kerr medium. 
We obtain approximate expressions for the time evolution operator as well as for the number 
operator in the Heisenberg re\-pre\-sentation.
We have found that the generation of photons from quantum vacuum is strongly a\-ffec\-ted by 
the presence of the Kerr medium, sharing physical features with the case of two two-level 
atoms in a cavity with osci\-llating walls. The nonlinear medium produces a saturation effect 
in the  photon generation which shows strong oscillations whose frequency increases with the
intensity of the nonlinear medium. 
We expect that these results could be relevant for any 
experimental study involving the dynamical Casimir effect that is willing to incorporate
Kerr nonlinearities.
%observed experimentally.

\end{abstract}


%\end{frontmatter}

%\linenumbers

\maketitle

\section{Introduction}
In 1970
it was predicted that in a one-dimensional electromagnetic
cavity made with perfectly conducting plane mirrors, one of which oscillates rapidly,
the generation of real photons from vacuum is possible~\cite{Moore1}. 
This phenomenon is known today as the dynamical Casimir effect (DCE) and it is considered
as a direct proof of the existence of quantum vacuum fluctuations of 
the electromagnetic field~\cite{RevModPhysFNori}.
Since then, a wide variety of theoretical works about the DCE
have been done~\cite{DodonovReview}.
For example, the generation and detection of vacuum photons has been 
studied in three-dimensional cavities~\cite{Dodonov1995126} and in 
time-dependent dielectric media~\cite{Klimov1}. However, experimental
realization of the effect has only been achieved recently~\cite{CasimirNature}
through the architecture of superconducting quantum circuits
where the effective length of the cavity (resonator) is rapidly
modulated \cite{Lahteenmaki12032013}.
Applications of the DCE as a means to create highly entangled states using
quantum circuits have been proposed in~\cite{Solano1}.
\medskip\\
Another well known nonlinear quantum optical phenomenon 
is the Kerr effect~\cite{boyd2008nonlinear}, where the refraction index of a material is proportional to the light 
intensity~\cite{fox2010optical}. 
It is usually used to generate non-classical states of light~\cite{Yurke1}. 
A typical situation is when a coherent state goes through a nonlinear Kerr
medium and evolves into a macroscopically distinguishable quantum 
superposition  of multiple coherent states, known as a {multi component} Schr\"odinger cat
state~\cite{gerry2005introductory}.
Experimental realization at the level of single-photon Kerr effect in 
three dimensional quantum circuits architecture was obtained in~\cite{KerrNature}.
\medskip\\
It has been shown that
Kerr nonlinearities should be taken into account for a realistic description	
of the DCE \cite{de2015microscopic}.
The motivation of the present work is to study 
how is the generation of vacuum photons modified due to the presence of a nonlinear Kerr medium.
This is an interesting question which lies at the heart of this contribution.
In order to gain some physical insight related to this question 
we have structured the paper as follows:
In section~\ref{Theory} we briefly review the theory related to the DCE and the
Kerr effects separately and present some known results relevant to this work.
In section~\ref{cavplusker}
we propose a Hamiltonian $H(t)$ that incorporates both effects and after several approximations we obtain a
Hamiltonian $\tilde \mathcal{H}_I$ that can be handled with algebraic methods whose time evolution operator 
$U_{\tilde \mathcal{H}_I}$ can be written in a product form.
In section~\ref{vacumphotons} the number operator is transformed into the Heisenberg representation
and the average number of photons from the vacuum is obtained. 
In order to justify the approximations made to get the interaction Hamiltonian $\tilde \mathcal{H}_I$,
 we also evaluate numerically the average number of photons generated from the vacuum state using 
 the full Hamiltonian $H(t)$ and confront our approximate results with the numerically converged.
Finally, in section \ref{conclusions} we give our conclusions.
\section{Theory}\label{Theory}
The simplest effective Hamiltonian that describes the DCE
in absence of dissipation is given by (with $\hbar=1$)
\cite{Dodonov1995126,Lahteenmaki12032013,Law1,DodonovOneAtom,DodonovTwoAtoms}
\begin{equation}\label{cavidad-vacia}
H_c=\omega(t)a^\dagger a+i\chi(t)(a^{\dagger 2}-a^2),
\	\	\	\chi(t)=\frac{1}{4\omega(t)}\frac{d\omega(t)}{dt}.
\end{equation}
where $a$, $a^\dagger$ are the usual bosonic field operators and
$\omega(t)$ is the time-dependent instantaneous frequency of the cavity.
The above Hamiltonian represents a single resonant electromagnetic field mode
within a cavity with perfectly conducting plane mirrors. 
In the DCE one of the mirrors is fixed while the other is oscillating  with
small amplitude. 
The oscillation frequency is set as twice the selected cavity mode 
$\omega_0$ (when the cavity has a length $L_0$) and 
the explicit temporal dependence in the frequency is chosen as:
$\omega(t)=\omega_0[1+\epsilon\sin(2\omega_0 t)]$~\cite{DodonovOneAtom}.
Here, $\epsilon$ is a small modulation amplitude and $2\omega_0$ is
the modulation frequency. If the mirror does not oscillate with a 
frequency close to $2\omega_0$ an infinite sum of interaction terms
appear on $H_c$  and though the system is complicated,
the equations of motion can still be solved~\cite{Klimov1}.
\subsection{Empty cavity}
In the context of DCE  Eq.~(\ref{cavidad-vacia}) is known as
the empty ca\-vity Hamiltonian, and it has been shown that 
the generation of photons from the quantum va\-cuum 
grows exponentially with time~\cite{Dodonov1995126}. 
In order to see that, we notice that for $\epsilon\ll 1$, 
one can write $\omega(t)\approx \omega_0$ since the influence of the
modulation is relevant mainly for the squeezing coefficient
\cite{PhysRevLett.84.1882}. Then, $\chi(t)\approx(\epsilon\omega_0/2)\cos(2\omega_0t)$ and  we obtain 
$H_c\approx\omega_0 a^\dagger a+
(i\epsilon\omega_0/2)\cos(2\omega_0t)(a^{\dagger 2}-a^2)$.
If we move to a reference frame generated by the unitary transformation 
$U_{1}=\exp(-i\omega_0ta^\dagger a)$ we get
$H_c^I =(i\epsilon\omega_0/4)(a^{\dagger 2}-a^2+a^{\dagger 2}e^{i4\omega_0t}-a^2e^{-i4\omega_0 t}) $.
In the previous equation, the last  two terms oscillate very rapidly and it is possible to make
use of the rotating wave approximation (RWA) and obtain a time-independent Hamiltonian:
${H}_c^I\approx (i\epsilon\omega_0/4)(a^{\dagger 2}-a^2)$,
whose time evolution operator is
$U_c^I=\exp[\frac{1}{2}r(a^{\dagger 2}-a^2)]$, which we can
identify as the squeezing operator with $r$=$\epsilon\omega_0t/2$
being the squeezing parameter.
The time evolution operator generates the following transformation in the field ope\-rators:
$a^\dagger(t) =a^\dagger \cosh(r)+a \sinh(r)$ and $a(t) =a^\dagger \sinh(r)+a \cosh(r)$.
We can now compute the average number of photons between the vacuum state at an arbitrary time $t$ as:
\begin{equation}\label{empty-cavity}
\langle 0|a^\dagger(t)a(t)|0\rangle=\sinh^2\left(\epsilon\omega_0t/2\right).
\end{equation}
For $\epsilon\omega_0t>1$ we can see the well known exponential growth
in the generation of vacuum photons~\cite{Klimov1}. This growth is 
a purely quantum manifestation of parametric amplification of vacuum fluctuations.
\subsection{Non-linear Kerr medium}
On the other hand, in Ref.~\cite{KerrNature} the authors present the experimental realization
of macros\-copically distinguishable superpositions of coherent states
using the Kerr Hamiltonian~\cite{Yurke1,walls2008quantum}:
\begin{equation}\label{kerr-hamiltonian}
H_K=\frac{K}{2}a^{\dagger 2}a^2.
\end{equation}
The parameter $K$ is the Kerr frequency shift per photon and
it is proportional to the third-order non-linear susceptibility~\cite{boyd2008nonlinear}.
One can show that the revival time for the Hamiltonian $H_K$
is given by $T_{K}=2\pi/K$~\cite{gerry2005introductory}.
The co\-rres\-pon\-ding evolution operator is
$U_K=\exp\left(-iKta^{\dagger 2}a^2/2\right)$.
Its action on a coherent state $|z\rangle$ is
$|z(t)\rangle=U_K|z\rangle=e^{-i(K/2)t(a^\dagger a)^2}|ze^{iKt/2}\rangle$. 
At a time $t=T_{K}/2$ the coherent state has evolved into
$|z(\pi/K)\rangle=e^{-i\pi/4}(|iz\rangle+i|-iz\rangle)/\sqrt{2}$.
This quantum state can be identified, except for a phase factor
as a Yurke-Stoler state \cite{Yurke1}, namely, a cat state.
Furthermore, the operator $U_K$
ge\-nerates the following transformation on the field operators:
$a^\dagger(t)$=${U_K^{\dagger}a^{\dagger}U_K}$=$a^\dagger e^{iKt(a^\dagger a)}$
and $a(t)$=${U_K^{\dagger}aU_K}$=$e^{-iKt(a^\dagger a)}a$.
It should be noted that the expectation value of the photon number operator remains
unchanged under the action of the Kerr Hamiltonian.
\section{Cavity with Kerr medium}\label{cavplusker}
Motivated by recent experiments related with the DCE~\cite{CasimirNature}  and
the Kerr effect~\cite{KerrNature} in superconducting quantum circuits,
we decided to study a system where both effects are present simultaneously.
Our proposal is based in a Hamiltonian  consisting of two parts:
the first one corresponds to a re\-sonant mode of the electromagnetic field within a 
cavity with oscillating walls and the second one to a Kerr
medium filling the cavity. The Hamiltonian is thus:
\begin{equation}\label{full-hamiltonian}
{H}(t)=\omega(t)a^\dagger a+i\chi(t)(a^{\dagger 2}-a^2)+\frac{K}{2}a^{\dagger 2}a^2.
\end{equation}
A similar Hamiltonian was proposed recently in~\cite{leon2015generation},
where time-dependent Kerr nonlinearities were analyzed.
It must be emphasised that the system re\-pre\-sented by Eq.~(\ref{full-hamiltonian})
is different from that of a one-mode electromagnetic field interacting with a movable mirror through the radiation pressure~\cite{Law2} 
(optomechanical system). There, the degrees of freedom of the
mirror are treated quantum mechanically, reducing the problem to a system of two coupled 
quantum oscillators. For completeness and illustration consider the Hamiltonian
describing such an interaction~\cite{KnightMirror}:
\begin{equation}\label{opto_mechanical}
H_{f-m}=\omega a^\dagger a+\nu b^\dagger b -\lambda a^\dagger a(b^\dagger +b),
\end{equation}
where $\nu$ ($\omega$) is the mirror (field) oscillation frequency  and  $b$, $b^\dagger$ 
($a$, $a^\dagger$) are the corresponding annihilation and creation operators. 
The coupling constant is  $\lambda=\frac{\omega}{L}\sqrt{\frac{\hbar}{2m\nu}}$, 
with $L$ the cavity length and $m$ the mass of the movable mirror.
Notice that in this system the frequencies $\nu$ and $\omega$ differ by many orders of magnitude.
The presence of the nonlinear term  $(a^\dagger a)^2$  in the co\-rres\-pon\-ding time evolution operator $U_{f-m}$
can lead to  the generation of non-classical states of light~\cite{ManciniOpto}.  
A confusion between the Hamiltonians (\ref{full-hamiltonian}) and
(\ref{opto_mechanical}) could arise, and one might expect that they describe the same dynamics. 
However, in the optomechanical system the nonlinear term  (Kerr medium) is only induced
and the squeezing term does not exist, making it difficult to study the DCE.
\medskip\\
Returning to Eq.~(\ref{full-hamiltonian}), we see that the Hamiltonian 
is  time-dependent and  the operators do not commute with each
other. This set of operators does not generate a Lie 
algebra and thus finding an exact solution for the Schr\"odinger equation 
is a real challenge. The system shows high algebraic complexity. 
In order to obtain {manageable} expressions, we apply appro\-ximations
{similar to those} used to get $H_c^I$. 
{Then, we obtain a time-independent Hamiltonian}
$\mathcal{H}=(i\epsilon\omega_0/4)(a^{\dagger 2}-a^2)+Ka^{\dagger 2}a^2/2$.
This Hamiltonian is well known in the context of non-linear quantum optics \cite{Gerry1987}
and it was proposed to study signatures of quantum chaos \cite{Milburn1990},
although {as far as we know} its general analytical solution is still missing. 
The corresponding time evolution operator is $U_{\mathcal{H}}=\exp\left(-i\mathcal{H}t\right)$. 
Unfortunately, this expression is not convenient because it can not be written in a product form.
Moreover, if we are looking for
$O(t)=U_{\mathcal{H}}^\dagger O U_{\mathcal{H}}$, being $O$ any physical
observable it would be very complicated to carry out this transformation.
Ideally, one would like to obtain a time evolution operator  written as a 
product of exponentials. 
In order to write the time evolution operator in a product form
we first go into the interaction picture representation generated by
the unitary transformation $U_2=\exp\left(-iKt(a^{\dagger}a)^2/2\right)$ which yields
\begin{equation}
\mathcal{H}_I= i\frac{\epsilon\omega_0}{4}\left(a^{\dagger 2}e^{i2Kt(a^{\dagger}a+1)}
-e^{-i2Kt(a^{\dagger}a+1)}a^2\right)-\frac{K}{2}a^{\dagger}a,
\end{equation}
where the identity $aF(a^\dagger a)=F(a^\dagger a+1)a$ was used and where
$F(a^\dagger a)$ is an arbitrary function of the number operator ($a^\dagger a$).
{Renaming the operators present in the Hamiltonian as}
\begin{eqnarray}\label{algebrasu11}
L_0=\frac{1}{2}\left(a^{\dagger}a+1/2\right),\	\
L_{+}(t)=\frac{1}{2}{a}^{\dagger 2}e^{i2Kt(a^{\dagger}a)},\	\ 
L_{-}(t)=\frac{1}{2}e^{-i2Kt(a^{\dagger} a)}{a}^2,
\end{eqnarray}
{and defining the function} $f(t)=ige^{i2Kt}$ with $g=\epsilon\omega_0/2$, 
we can rewrite the Hamiltonian as
$\mathcal{H}_I=-KL_0+f(t)L_+(t)+f^*(t)L_{-}(t)+K/4$.
Notice that the term $K/4$ commutes with all the other elements
and it will only gene\-rate a global phase $e^{-iKt/4}$ in the dynamics.
The remaining time-dependent operators satisfy the commutation relations:
\begin{equation}\label{rel-conmutacion}
[L_{-}(t),L_{+}(t)]=2L_0,\	\	\	\	[L_0,L_{\pm}(t)]=\pm L_{\pm}(t),
\end{equation}
\textcolor{black}{
which we can identify as those of the $su(1,1)$ Lie algebra~\cite{klimov2009a}. 
It is important to mention that it is possible to find a unitary transformation that
diagonalizes the Hamiltonian $\mathcal{H}_I$ and makes it time-independent. 
Consider a generalized displacement operator $D(\zeta,t)=\exp[\zeta L_+(t)-\zeta^* L_-(t)]$
with the constrictions $\tanh(2|\zeta|)=-2g/K$ and $\zeta=|\zeta|e^{i2Kt+i\pi/2}$.
Transforming $\mathcal{H}_I$ we get $D(\zeta,t)\mathcal{H}_ID^\dagger(\zeta,t)=-\sqrt{K^2-4g^2}L_0$.
Applying the standard decomposition formulas \cite{Ban} for the $su(1,1)$ algebra the
displacement operator can be factorized as.}
\textcolor{black}{
\begin{equation}
D(\zeta,t)=e^{\frac{\zeta}{|\zeta|}\tanh|\zeta|L_+(t)}e^{-2\ln\cosh|\zeta|L_0}
e^{\frac{-\zeta^*}{|\zeta|}\tanh|\zeta|L_-(t)}.
\end{equation} }
\textcolor{black}{
If we use the Fock states $|n\rangle$ and the definition operator of $L_0$
we arrive to the following identity}
\textcolor{black}{
\begin{eqnarray}
\mathcal{H}_ID^\dagger(\zeta,t)|n\rangle=
-\sqrt{(K/2)^2-g^2}\left(n+{1}/{2}\right)D^\dagger(\zeta,t)|n\rangle,
\end{eqnarray} } 
\textcolor{black}{with $n$ a non-negative integer. States of the form $|n\rangle_t=D^\dagger(\zeta,t)|n\rangle$ 
generate a complete basis in which the interaction picture Hamiltonian is diagonal.
It is interesting to notice that eigenvalues of $\mathcal{H}_I$ are equally spaced 
at any  time just as the eigenvalues of the so called Ermakov-Lewis invariant
of the quantum time-dependent harmonic oscillator \cite{Lewis}.}
\medskip\\
Due to the fact that
the Hamiltonian has been written in terms of the generators of the $su(1,1)$ Lie algebra~\cite{klimov2009a},
one could naively try to make use of the Wei-Norman theorem \cite{Norman};
which allows us to write  the time evolution operator as a pro\-duct of exponentials.
However, in order to use such a theorem, the algebra generators must be time-independent. 
From Eq. (\ref{algebrasu11}) we clearly see their explicit time dependence,
therefore, we cannot use it. 
However, for an infinitesimally short interval of time from $t$ to $t+\delta t$,
the evolution operator is \cite{gerry2005introductory}
\begin{eqnarray}
U_{\mathcal{H}_I}&\cong & \exp[-i\mathcal{H}_I\delta t],\nonumber\\
&=&\exp\left[i\delta t KL_0-i\delta tf(t)L_+(t)-i\delta tf(t)^*L_{-}(t)\right],\nonumber\\
&=& \exp[g_1(t)L_0]\exp[g_2(t)L_+(t)]\exp[g_3(t)L_{-}(t)],\nonumber
\end{eqnarray}
with the complex time-dependent functions $g_n(t)$ to be easily determined \cite{Ban}. 
For a finite interval of time, say from $0$ to $T$, the time
evolution operator can be written as
\begin{equation}
U_{\mathcal{H}_I}=\lim_{\delta t\rightarrow 0}\mathcal{T}\prod_{l=0}^{T/\delta t}
e^{g_1(t_l)L_0}e^{g_2(t_l)L_+(t_l)}e^{g_3(t_l)L_{-}(t_l)}.
\end{equation}
$\mathcal{T}$  is the usual time ordering operator and $t_l=l\delta t$. 
Notice that for each time $t_l$, the evolution operator factorizes as a 
product of exponentials.
Without the exponential term of the number operator
we would have the squeezing operator, but this is not  
the case and we are dealing here with a modified squeezing operator.
The total time evolution operator has the form:
$U(t)\approx U_1U_{\mathcal{H}}=e^{-iKt/4}U_1U_2U_{\mathcal{H}_I}$.
\medskip\\	
It turns out that our problem is to find  an efficient way to apply
$U_{\mathcal{H}_I}$ to an arbitrary initial state. This is still too complicated and we will 
require another approximation in order to obtain closed analytical expressions
which can be compared with the exact numerical results.
The approximation consists in making $e^{\pm i2Kt(a^\dagger a)}\approx1$ in the 
generators of the $su(1,1)$ Lie algebra of Eq. (\ref{algebrasu11}).
This approximation can be justified when the average value of the number operator is much
less than one which happens at very early times when the photon generation  is star\-ting.
We will study the early evolution of the system for times much smaller than the revival time.
The latter is generally much larger than the classical period~\cite{Robinett20041}.
Within this approximation, the operators in the  interaction Hamiltonian become time-independent:
\begin{equation}\label{third-approx}
\mathcal{H}_I\approx\tilde{\mathcal{H}}_I=-KL_0+f(t)L_++f^*(t)L_{-}.
\end{equation}
Before proceeding with the calculations, there are several points we want to stress.
The information of the Kerr  medium is still encoded in Eq.~(\ref{third-approx}) due to the
$K$ dependence of the function $f(t)$ and also in the operator $U_2$.
The algebraic structure of $\tilde{\mathcal{H}}_I$ is equal to that of a degenerate
parametric amplifier~\cite{walls2008quantum}. 
It is time-dependent and does not commute with itself at different times, so that we cannot write it as
$U_{\tilde{\mathcal{H}}_I}=\exp(-i\int \tilde{\mathcal{H}}_Idt)$.
Fortunately, it consists of a set of time-independent operators forming
the $su(1,1)$ Lie algebra. At this point we can apply the Wei-Norman
theorem~\cite{Norman} and write the time evolution operator as a product of exponentials:
\begin{equation}
U_{\tilde{\mathcal{H}}_I}=e^{\alpha(t)L_+}e^{\beta(t)L_{0}}e^{\gamma(t)L_{-}},
\end{equation}
where the complex time-dependent functions $\gamma(t)$, $\beta(t)$, $\alpha(t)$ satisfy the following set of
coupled non-linear ordinary differential equations 
\cite{klimov2009a} 
\begin{eqnarray}
\dot{\alpha}(t)	 &=&-i\left[f(t)-K\alpha(t)+f^*(t)\alpha^2(t)\right],\nonumber\\
\dot{\beta}(t) 	 &=&-i\left[-K+2f^*(t)\alpha(t)\right],\nonumber\\
\dot{\gamma}(t) &=&-if^*(t)e^{\beta(t)}.
\end{eqnarray}
The abovementioned equations are obtained using the time-dependent 
Schr\"o\-dinger equation in the interaction representation.
The equation for  $\alpha(t)$ is the well known Ricatti di\-ffe\-rential equation, which in this case has
an exact analytical solution for the given initial conditions $\alpha(0)$=$\beta(0)$=$\gamma(0)$=$0$.
Therefore, functions $\beta(t)$ and $\gamma(t)$ can be 
easily obtained by direct integration:
\begin{eqnarray}\label{differential-equations}
\alpha(t)&=&\frac{ge^{i2Kt}\sinh(t\eta)}{\eta\cosh(t\eta)+
i(K/2)\sinh(t\eta)},\nonumber\\
\beta(t)&=& i2Kt+2\ln(\eta)-2\ln\left[ \eta\cosh(t\eta)+
i(K/2)\sinh(t\eta)\right],\nonumber\\
\gamma(t)&=&\frac{-g\sinh(t\eta)}{\eta\cosh(t\eta)+
i(K/2)\sinh(t\eta)},
\end{eqnarray}
 where we have defined $\eta=\sqrt{g^2-(K/2)^2}$.
If $K/2>g$ the hyperbolic functions in Eq.~(\ref{differential-equations}) 
have to be replaced with their trigonometric counterparts and $\eta\rightarrow\tilde{\eta}=\sqrt{(K/2)^2-g^2}$.
In this way, the approximate time evolution operator for the total system is:
$U(t)$ $\approx$ $e^{-iKt/4}U_1U_2U_{\tilde{\mathcal{H}}_I}$=$\mathcal{U}(t)$.
Explicit substitution of each product in terms of the field operators leads to:
\begin{eqnarray}\label{evolution-operator}
\mathcal{U}(t)&=&\exp\left[\frac{\beta(t)}{4}-\frac{iKt}{4}\right]
\exp\left[-i\omega_0ta^\dagger a-\frac{iKt}{2}(a^\dagger a)^2\right]
\nonumber\\&\times&
\exp\left[{\frac{\alpha(t)}{2}a^{\dagger 2}}\right]
\exp\left[{\frac{\beta(t)}{2}a^{\dagger}a}\right]
\exp\left[{\frac{\gamma(t)}{2}a^2}\right].
\end{eqnarray}
Eq.~(\ref{evolution-operator}) is one of the main results of 
this work, providing a closed analytical form for the time evolution operator
of the system at short times.
Notice that $\mathcal{U}(t)$ is formed by a squeezing part, generated by 
the terms $a^{\dagger 2}$, $a^2$ and two other terms 
that generate linear $a^{\dagger}a$ and non-linear $(a^\dagger a)^2$ evolution.
%\textcolor{red}{Due to $\mathcal{U}(t)$ is a product of exponentials, apply it to 
%any initial state of interest is an straightforward task. In a full and explicit
%manner one can elucidate the quantum evolution of the photon state.}
\section{Generation of vacuum photons}\label{vacumphotons}
Now we are in a position from which we can analyze
how the photon generation from quantum vacuum is affected when a Kerr medium is present. 
For such a task we need the number operator, $N=a^\dagger a$ in the Heisenberg representation:
\begin{equation}
N(t)=\mathcal{U}^\dagger (t) a^\dagger a\mathcal{U}(t)=\Phi_1a^{\dagger 2}+
\Phi_2a^\dagger a+\Phi_3a^2+\Phi_4,
\end{equation}
where we have defined
\begin{eqnarray}
&&\Phi_1=\alpha(t)e^{-\beta(t)},\	\	\	\	\	\	\	
\	\	\	\	\	\	\	
\Phi_3=\alpha(t)\gamma(t)^2e^{-\beta(t)}-\gamma(t),\nonumber\\
&&\Phi_2=1-2\alpha(t)\gamma(t)e^{-\beta(t)},\	\	\
\Phi_4=-\alpha(t)\gamma(t)e^{-\beta(t)}.
\end{eqnarray}
Although the operator $N(t)$ does not look Hermitian, actually
it is, since it comes from a transformation given by the operator $\mathcal{U}(t)$ 
which is unitary by construction through the Wei-Norman theorem. 
Due to the approximations carried out, one could be tempted to doubt about the Hermiticity of $N(t)$. 
However, these transformations were performed on the Hamiltonian 
operator and they never compromised its Hermiticity.
In fact, direct substitution of  
$\alpha(t)$, $\beta(t)$ and $\gamma(t)$ yield  $\Phi_1=\Phi_3^*$
with $\Phi_2$, $\Phi_4$ real functions so that $N(t)$ is Hermitian. 
\medskip\\
Having obtained the operator $N(t)$, it is a simple matter to calculate 
the ave\-rage number of vacuum photons
$\langle N\rangle_0=\langle 0|N(t)|0\rangle=\Phi_4$:
\begin{equation}\label{kerr-cavity}
\langle N\rangle_0=\frac{g^2}{g^2-({K}/{2})^2}
\sinh^2\left(t\sqrt{g^2-\left({K}/{2}\right)^2}\right).
\end{equation}
From Eq.~(\ref{kerr-cavity}) we can identify three distinct regimes:
$i)$ for $K/2<g$, there is an exponential growth in the average number of vacuum photons. 
In fact, this formula is a generalization of Eq.~(\ref{empty-cavity}). 
Obviously, both equations coincide when $K=0$.
$ii)$ For  $K/2>g$ the equation takes its trigonometric form and $\langle N\rangle_0$
will have strong oscillations, being zero at times $t=2m\pi/\tilde{\eta}$ with $m$ a positive integer.
$iii)$ For $K/2\rightarrow g$, $\langle N\rangle_0$ grows as $t^2$.
\textcolor{black}{The later case is of special interest, because the DCE and Kerr
effect contribute equally into the dynamics and the photon production grows monotonically in a 
slower manner as compared with the $i)$ case. This quadratic function is precisely 
the separatix between the exponential growth and the oscillatory behavior.}
\begin{figure}[h]
\begin{center}
\includegraphics[width=9cm, height=6cm]{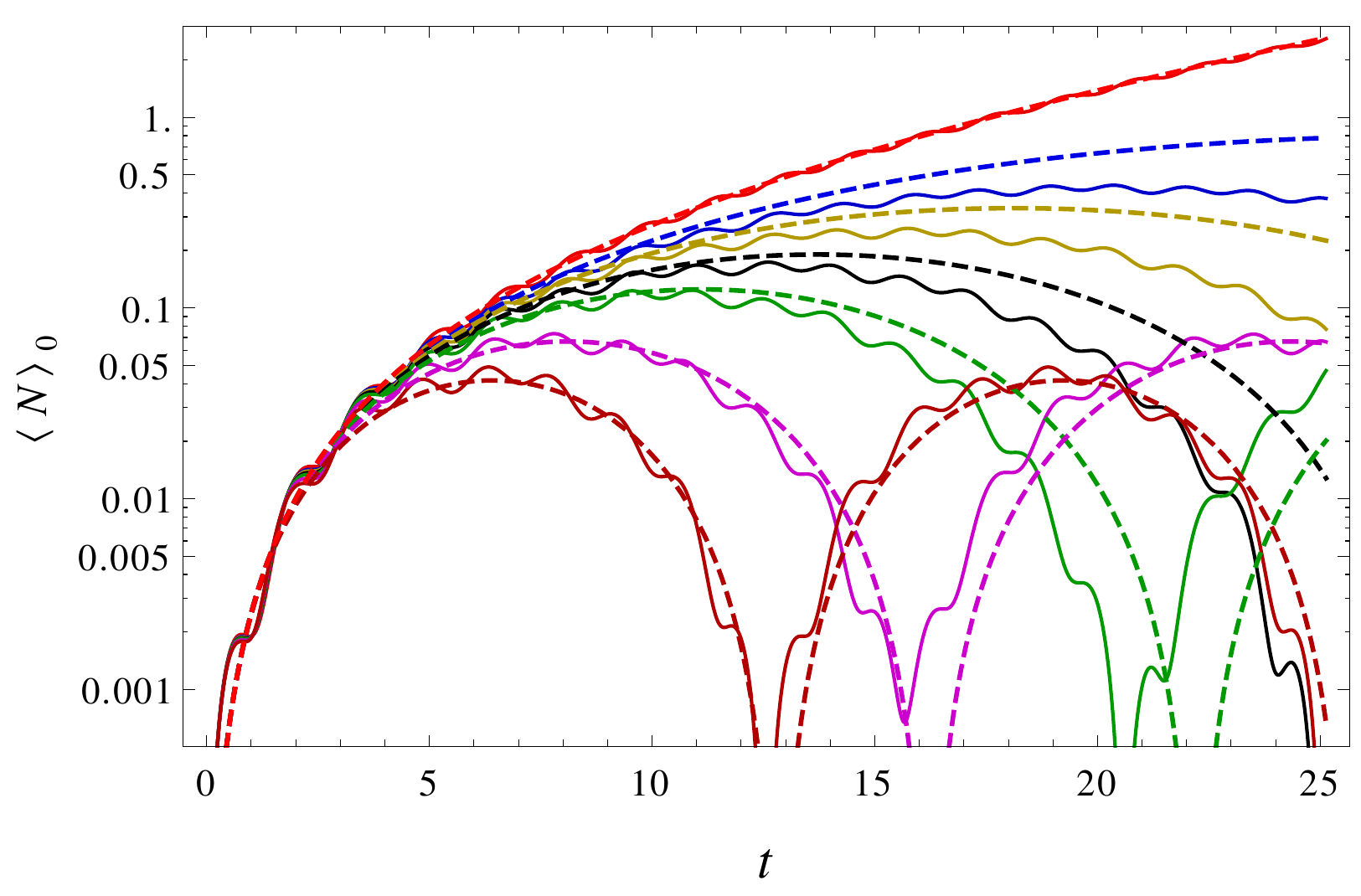}
\caption{
Generation of photons from quantum vacuum.
Solid (dashed) lines correspond to the numerical (analytical) solutions. 
We have set $\omega_0=1$, $\epsilon=0.1$ and $K=0$ 
(light red), $0.15$ (blue), $0.2$ (yellow), $0.25$ (black), $0.3$ (green),
$0.4$ (magenta) and $0.5$ (dark red).}
\label{photon_vacum}
\end{center}
\end{figure}\\
In Fig. \ref{photon_vacum} we show $\langle N\rangle_0$ as a function of time for the
analytical results given by Eq.~(\ref{kerr-cavity}) and the converged numerical calculation 
using the full Hamiltonian of Eq.~(\ref{full-hamiltonian}).
We have used $\omega_0=1.0$, $\epsilon=0.1$ and different relevant values for $K$.
As a re\-fe\-rence, the light-red line corresponds to the case when $K=0$, 
showing the expected exponential growth  according to  Eq.~(\ref{empty-cavity}).
However, when $K\neq 0$ a rapid decrease in the
vacuum photon number accompanied by sharp oscillations is observed, as
predicted by Eq.~(\ref{kerr-cavity}). Physically, these oscillations
could be due to a saturation effect of the non-linear material~\cite{boyd2008nonlinear}. 
The numerical solution also shows small oscillations absent in the
analytical solution; these were eliminated by the RWA. The smallest revival time
used was $T_{0.5}=4\pi\approx 12$, which corresponds to the 
largest value of $K$ used in the figure. The other revival times are obviously larger.
Clearly, there is a good agreement between analytical and numerical calculations 
for times $t\ll T_{K}$ and $\langle N\rangle_0\ll1$. 
After this time, it is natural to see substantial differences.
\begin{figure}[h]
\begin{center}
\includegraphics[width=8.5cm, height=6.3cm]{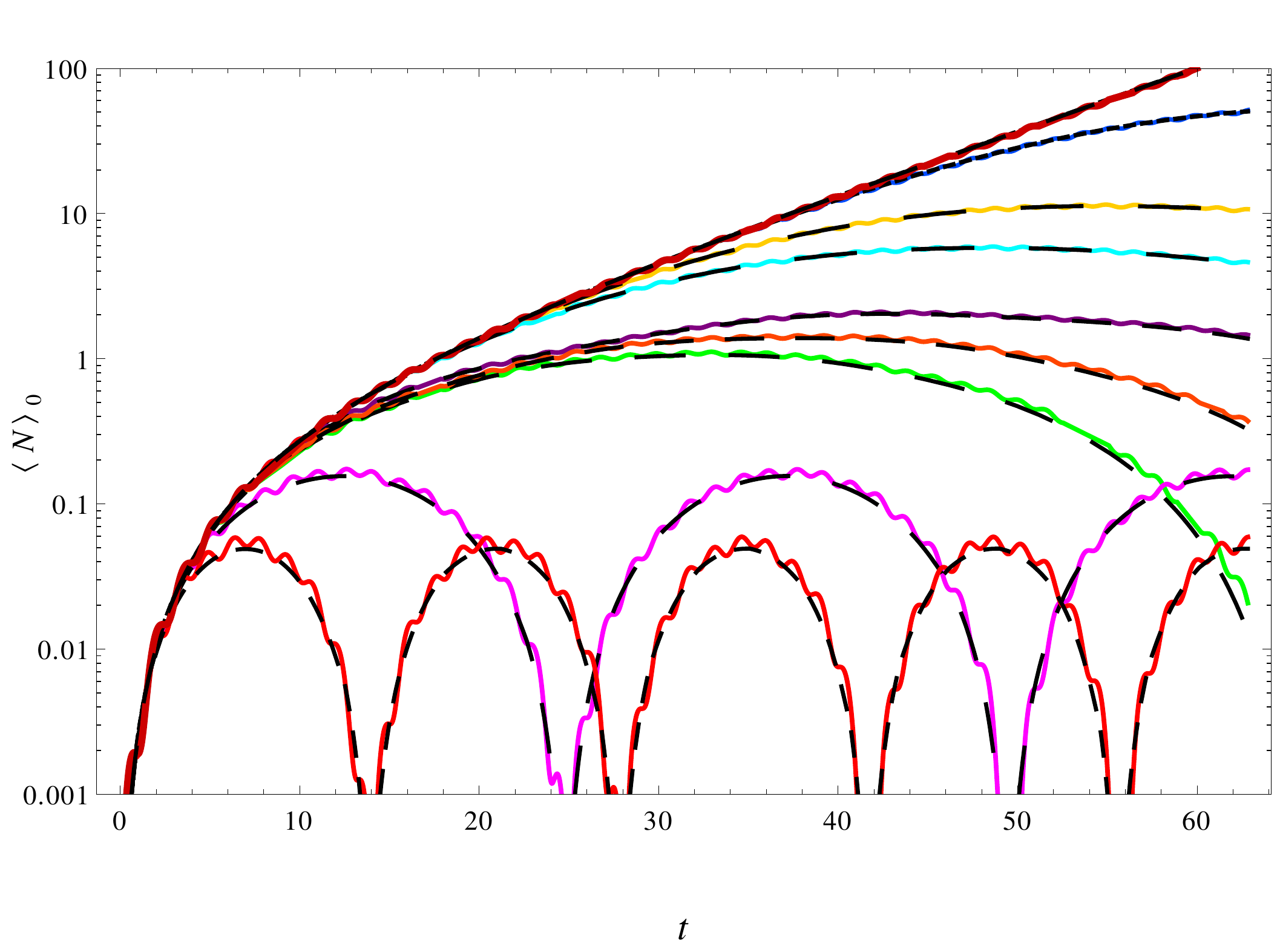}
\caption{
Generation of photons from quantum vacuum.
Solid (black-dashed) lines correspond to the full numerical solutions without (with) the RWA. 
We have set $\omega_0=1$, $\epsilon=0.1$ and $K=0$ 
(dark-red), $0.001$ (blue), $0.005$ (yellow), $0.01$ (cyan), $0.05$ (dark-purple), 
$0.07$ (orange),  $0.085$ (green), $0.25$ (magenta) and $0.45$ (light-red).}
\label{photon_vacum_num}
\end{center}
\end{figure}\\
\textcolor{black}{In order to see what happens with the dynamics at longer times and to explore 
the large average photon number regime, we performed a purely numerical
calculation making the evolution using}
the full Hamiltonian given by Eq. (\ref{full-hamiltonian}).
\textcolor{black}{The results are shown in solid lines in Fig. \ref{photon_vacum_num}.}
The dashed lines correspond also to the numerical solution but using the RWA, showing
again the absence of the small oscillations.
The overall picture is very similar to {that of} Fig. \ref{photon_vacum}, 
\textcolor{black}{however, there are important differences. In the numerical solution, for any value of $\epsilon$
and provided that Kerr coefficient $K$ is different from zero, the oscillatory 
behavior will always be present. When $K$ is extremely small this behavior will appear
in a very long time scale. This is the reason why an apparent infinite expo\-nen\-tial 
photon growth at short times is displayed in the $i)$ case.}
\medskip\\		
In the regime where the average photon number is large we do not expect our approximate solution
to match the numerical one. However,  we can reproduce the
qualitative behavior in this regime by tuning the $K$ parameter to a slightly larger value 
in Eq. (\ref{kerr-cavity}).
The numerical analysis suggests the impossibility  of the asymptotic photon growth. This is
the main difference with the result of Eq. (\ref{kerr-cavity}).
In fact, it was proved in \cite{de2015microscopic}
that the photon generation is limited
for a time-independent Hamiltonian version of Eq. (\ref{full-hamiltonian}).
\medskip\\
It should be mentioned that Eq.~(\ref{kerr-cavity}) was
recently obtained in Ref.~\cite{DodonovEcGen} but in a different context. 
There, the authors considered the empty cavity Hamiltonian in Eq.~(\ref{cavidad-vacia})
and a small shift $\kappa\ll 1$ in the mirror oscillation frequency.
Similar behaviors of $\langle N\rangle_0$ are presented in 
\cite{DodonovTwoAtoms}, where two two-levels atoms (TLA) within a
cavity with oscillating walls are used as photon detectors,
showing that $\langle N\rangle_0$ is also $\ll 1$.
We can relate their results with ours and infer that the two TLA 
might behave like an intensity-dependent refraction index. It is in analogy to the Jaynes-Cummings model, 
where the state of the field, initially in a coherent state, evolves into distinguishable
quantum superpositions like the ones generated by a Kerr medium~\cite{haroche2006exploring}.
\medskip\\
\textcolor{black}{As a final remark, it would be interesting to consider the open 
dynamics of the model through a phenomenological Lindblad master equation. For instance, 
introducing dissipation in the principal cavity mode in order to investigate its effect of photon
production. Unfortunately, this approach is not well justified in the present case as it does
not include the nonstationary solution of the system Hamiltonian.  The ideal treatment for
the open system dynamics involve a formal microscopic derivation of the master equation
in which the exact eigenstates of the Hamiltonian can be used to construct the Lindblad
jump operators. This approach is out of the scope of the present work.}
%\newpage
\section{Conclusions}\label{conclusions}
We have studied the simplest form of a Hamiltonian that describes
the DCE and Kerr effects simultaneously.
An approximate time evolution operator for the whole
system was obtained. The evolution operator could be written as a product of exponentials
containing squeezing, linear and non-linear evolution. A closed analytical expression
for $\langle N\rangle_0$ was obtained and 
when we confronted the approximate results with converged numerical calculations we found good 
agreement for times shorter than the revival time and $\langle N\rangle_0\ll 1$.
We found that the vacuum photon generation, which initially has an exponential growth, 
can exhibit a rapid decrease with strong oscillations due to the presence of the Kerr medium.
This behavior becomes steep when the parameter $K$ increases, 
and is present at any time scale, short or large and also when $\langle N\rangle_0$ is much larger than one.
We have compared our results with a previous work establishing a link between 
two different systems. We found {that $\langle N\rangle_0$ shows a similar conduct for a Kerr
medium and two TLA  within a cavity with oscillating walls~\cite{DodonovTwoAtoms}.} 
From these results we infer that a system of two 
TLA can act as a non-linear medium with an intensity-dependent refraction index.
For future work, it would be interesting to see the time
evolution for the vacuum state in phase space representation, for instance,
by means of the Wigner function. Perhaps it could lead to the formation
of quantum superpositions coming from the quantum vacuum, a ``{\it vacuum cat state}".
\section*{Acknowledgments}
We thank M. A. Bastarrachea for his useful comments which improved the manuscript.
We acknowledge partial support from CONACyT through project 166961 and DGAPA-UNAM project IN108413.
RRA and CGG would like to express their gratitude to CONACyT for financial support under scholarships No.
379732 and No. 385108.

%\bibliography{casimir_dinamico}
%\bibliographystyle{ieeetr}
%\bibliographystyle{elsarticle-num}

\end{document}